\documentclass[epj,nopacs]{svjour}
\usepackage{graphicx}
\usepackage{epsfig}

\newcommand{\qslash}{q\!\!\!/}
\newcommand{\m}{\tiny{-}}
\newcommand{\p}{\tiny{+}}

\newcommand{\plaata}[4]{\raisebox{#2pt}{
\epsfig{figure=./#1.eps,
width=#3cm,height=#4cm}}}
\newcommand{\psb}{\bar{\psi}}

\begin{document}

\title{Cross sections for multi-particle final states 
       at a linear collider}  

\author{T.~Gleisberg \inst{1}
        \thanks{email: tanju@theory.phy.tu-dresden.de}          
        \and 
        F.~Krauss \inst{1,2} 
        \thanks{email: Frank.Krauss@cern.ch}
        \and 
        C.~G.~Papadopoulos \inst{2,3}
        \thanks{email: Costas.Papadopoulos@cern.ch}
        \and 
        A.~Sch{\"a}licke \inst{1} 
        \thanks{email: dreas@theory.phy.tu-dresden.de}
        \and 
        S.~Schumann \inst{1}
        \thanks{email: steffen@theory.phy.tu-dresden.de}
}

\institute{
           Institut f{\"u}r Theoretische Physik, TU Dresden, 
           01062 Dresden, Germany
   \and 
           Theory Division, CERN, 1211 Geneva 23, Switzerland
           \and 
           Institute of Nuclear Physics, NCSR Demokritos, 15310 Athens, 
           Greece
}

\abstract{
In this paper total cross sections for signals and backgrounds 
of top- and Higgs-production channels in $e^+e^-$ collisions
at a future linear collider are presented. All channels considered 
are characterized by the emergence of six-particle final states. 
The calculation takes into account the full set of tree-level 
amplitudes in each process. Two multi-purpose parton level 
generators, {\tt HELAC/PHEGAS} \cite{Kanaki:2000ey,Papadopoulos:2000tt} 
and {\tt AMEGIC++} \cite{Krauss:2001iv} are used, their results 
are found to be in perfect agreement.}

\date{November 21$^{\rm st}$ 2003}
\headnote{\small\rm CERN-TH/2003-282}
\maketitle

\section{Introduction}
\label{intro}
Six-particle final states constitute the signature for many
processes that will be studied at the precision level at a future
$e^+e^-$ linear collider. Important channels include the production
and decay of top quark pairs and -- if existent -- of one or more
Higgs bosons, the latter process allowing a test of the structure
of the Higgs potential. Furthermore, if no evidence for a Higgs 
boson was found at the LHC, the study of quartic gauge boson couplings
is mandatory in order to understand alternative scenarios of
electroweak symmetry breaking. Leaving the framework of the
Standard Model (SM) the production of, say, chargino pairs in the
Minimal Supersymmetric Standard Model (MSSM) will lead to six-particle 
final states as well. To understand these processes at the
precision level, i.e. at the order of a few per cent, it is mandatory 
to supplement typical approaches such as the narrow-width approximation,
with corresponding calculations through full amplitudes, and to 
quantify the effect of non-resonant contributions. Obviously, for 
hadronic final states, a full QCD calculation is unavoidable. 

\noindent
Such investigations, however, are a formidable calculational 
task that cannot be handled without dedicated computer programs.
Two major difficulties make these necessary:
\begin{enumerate}
\item Including the full SM for the production of a six-particle final
      state often leads to having to handle a large numbers of diagrams.
      As an illustrative example of this problem, take the process
      $e^+e^-\to e^+e^-e^+e^-e^+e^-$, which results in 13896 Feynman
      diagrams. Obviously, the common textbook method of  
      squaring the diagrams, by employing completeness relations for 
      the external particles and evaluating the traces, is not a very
      efficient way to calculate the matrix element squared.
\item Apart from the treatment of an enormous number of diagrams,
      growing roughly factorially with the number of external legs,
      the integration over the phase space of the outgoing particles
      becomes a tedious task. The high dimensionality, $3n-4$,  
      for $n$ final-state particles necessitates the use of Monte Carlo
      methods. To achieve convergence of the Monte Carlo procedure 
      process- and cut-dependent phase-space mappings are required
      that tame wildly fluctuating integrands, which are due to nearly 
      on-shell propagators. A benefit of Monte Carlo methods, if carefully
      implemented, is that not only total cross sections but also
      distributions including all possible types of kinematical cuts
      can be calculated on an equal footing.
\end{enumerate}

\noindent
In the past years, different types of parton level generators have
been constructed. They can be crudely characterized as either
specialized or multi-purpose generators. 

\noindent
Usually, the former ones contain explicit matrix elements and phase-space 
mappings for specific classes of processes with specific 
assumptions. These matrix elements were constructed before 
outside the respective program, and this feature also allows for 
instance to implement non-universal higher order corrections in a 
controlled way. Often, the phase space mappings can be optimized
before as well. 

\noindent
Examples for such programs dealing with some of the processes 
discussed here are {\tt LUSIFER} \cite{Dittmaier:2002ap} and 
{\tt eett6f} \cite{Kolodziej:2002ev}. Both are constrained to fermions 
in the final state; in the case of {\tt LUSIFER} these fermions are 
bound to be massless, whereas {\tt eett6f} specializes in top quark pair 
production channels where the outgoing fermions might be massive, 
but electrons are disallowed in the final state. Both programs use 
versions of the adaptive multichannel method in the spirit of
\cite{Berends:1994pv} for their integration. A further dedicated
program using the multichannel importance sampling is {\tt SIXFAP} 
\cite{Gangemi:1998vc}. It provides the electroweak contributions
for a large set of six-fermion final states, taking into account 
possible non-zero fermion masses.   

\noindent
In contrast to specialized programs, multipurpose codes generate
both the matrix elements and the phase space mappings with or without
some intervention by the user. Apart from the programs used in this
paper, examples of these types of programs are {\tt O'Mega/Whizard} 
\cite{Moretti:2001zz,Kilian:2001qz} and {\tt MadGraph/MadEvent} 
\cite{Stelzer:1994ta,Maltoni:2002qb}. In the first package, {\tt O'Mega} 
\cite{Moretti:2001zz} relies on a version of the {\tt alpha} algorithm 
\cite{Caravaglios:1995cd}. However, in the 
present version of {\tt O'Mega}, full QCD has not been implemented yet. 
The integration of the resulting matrix elements is achieved through 
{\tt Whizard} \cite{Kilian:2001qz}, which constructs phase-space mappings 
automatically and integrates them with the {\tt VAMP}-algorithm 
\cite{Ohl:1998jn}. In fact, {\tt Whizard} can also be interfaced with 
other matrix element generators and it can be used to generate 
unweighted, single events. In contrast, {\tt MadGraph/MadEvent} 
generates all Feynman diagrams for a process under consideration and
then passes the information to the {\tt HELAS} package 
\cite{Murayama:1992gi} for the translation into corresponding helicity 
amplitudes. The integration channels are constructed automatically,
and a new version of the adaptive multichannel method described in
\cite{Maltoni:2002qb} is employed for the actual integration and the
generation of unweighted events. 

\noindent
This paper deals with the calculation of total cross sections for many 
relevant processes at a future $e^+e^-$ collider with two different,
independent packages, namely {\tt HELAC/PHEGAS} and {\tt AMEGIC++}.
Similar to the comparison of four-fermion generators at the LEP2
Monte Carlo Workshop \cite{Grunewald:2000ju}, a detailed study and
mutual benchmarking of tools for six- and eight-particle final states
at a future linear collider has been initiated in the framework of 
the extended ECFA/DESY study \cite{Aguilar-Saavedra:2001rg}. Here, 
a further step into this direction is reported. 

For the case of only massless final state particles, a similar comparison 
between the programs {\tt LUSIFER} and {\tt MADGRAPH}, the latter using
{\tt WHIZARD} for the phase-space integration has been presented in 
\cite{Dittmaier:2002ap}. In addition, results achieved by different 
generators for selected top quark pair production channels in the 
massless fermion approximation can be found in \cite{Dittmaier:2003sc}.

\noindent
The paper is organized as follows. In Sect. ~\ref{manHELAC}
and ~\ref{manAMEGIC} the relevant features of the two programs, {\tt HELAC/PHEGAS} 
and {\tt AMEGIC++}, are briefly reviewed. In  Sect.~\ref{numResults} the results are
presented and discussed. Conclusions are drawn in Sect.~\ref{Conclusion}. 

\section{The {\tt HELAC/PHEGAS} package}\label{manHELAC}
%
\subsection{Amplitude Computation: {\tt HELAC}}

\noindent
The traditional representation of the scattering amplitude in terms 
of Feynman graphs results in a computational cost that grows like 
the number of those graphs, therefore as $n!$, where $n$ is the
number of particles involved in the scattering process. 

\noindent
An alternative to the Feynman graph representation is provided 
by the Dyson-Schwinger approach. Dyson-Schwinger equations 
recursively express the $n$-point Green's functions
in terms of the $1$-,$2$-,$\ldots,(n-1)$-point functions.
For instance in QED these equations can be written as follows:
\begin{eqnarray}
\plaata{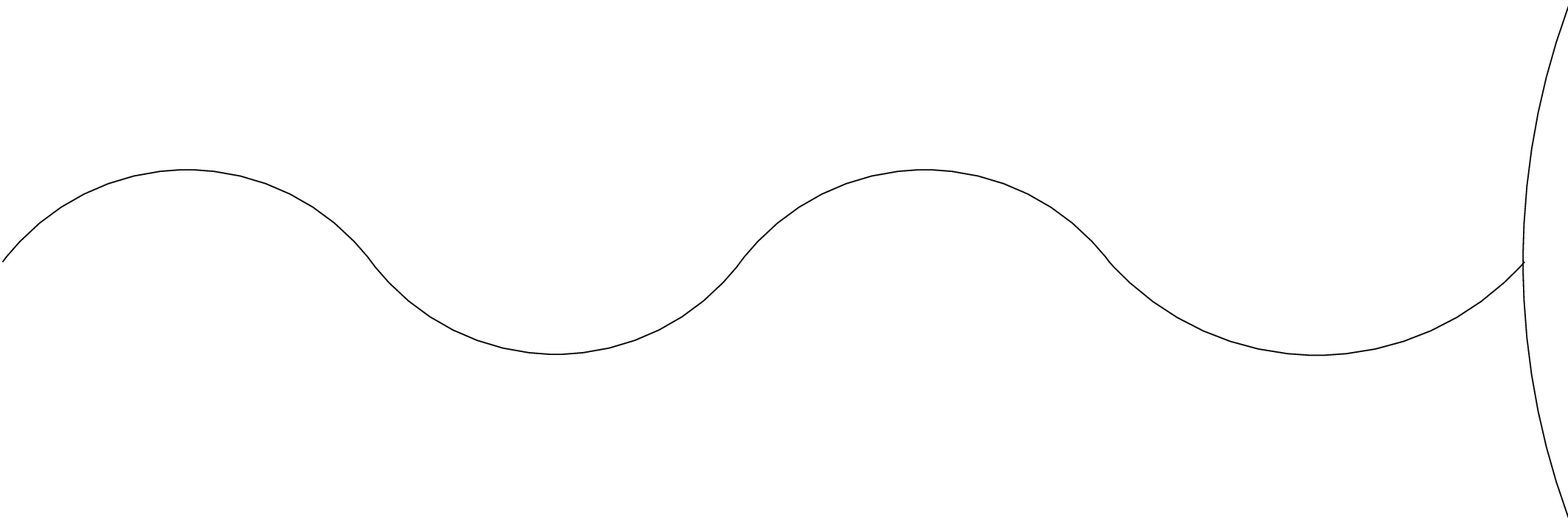}{-15}{1.8}{1}\;\;=\;\plaata{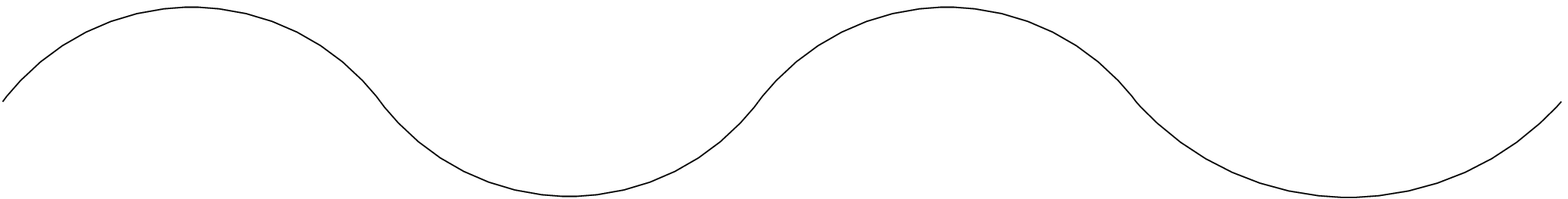}{0}{1.8}{0.2}\;
+\;\plaata{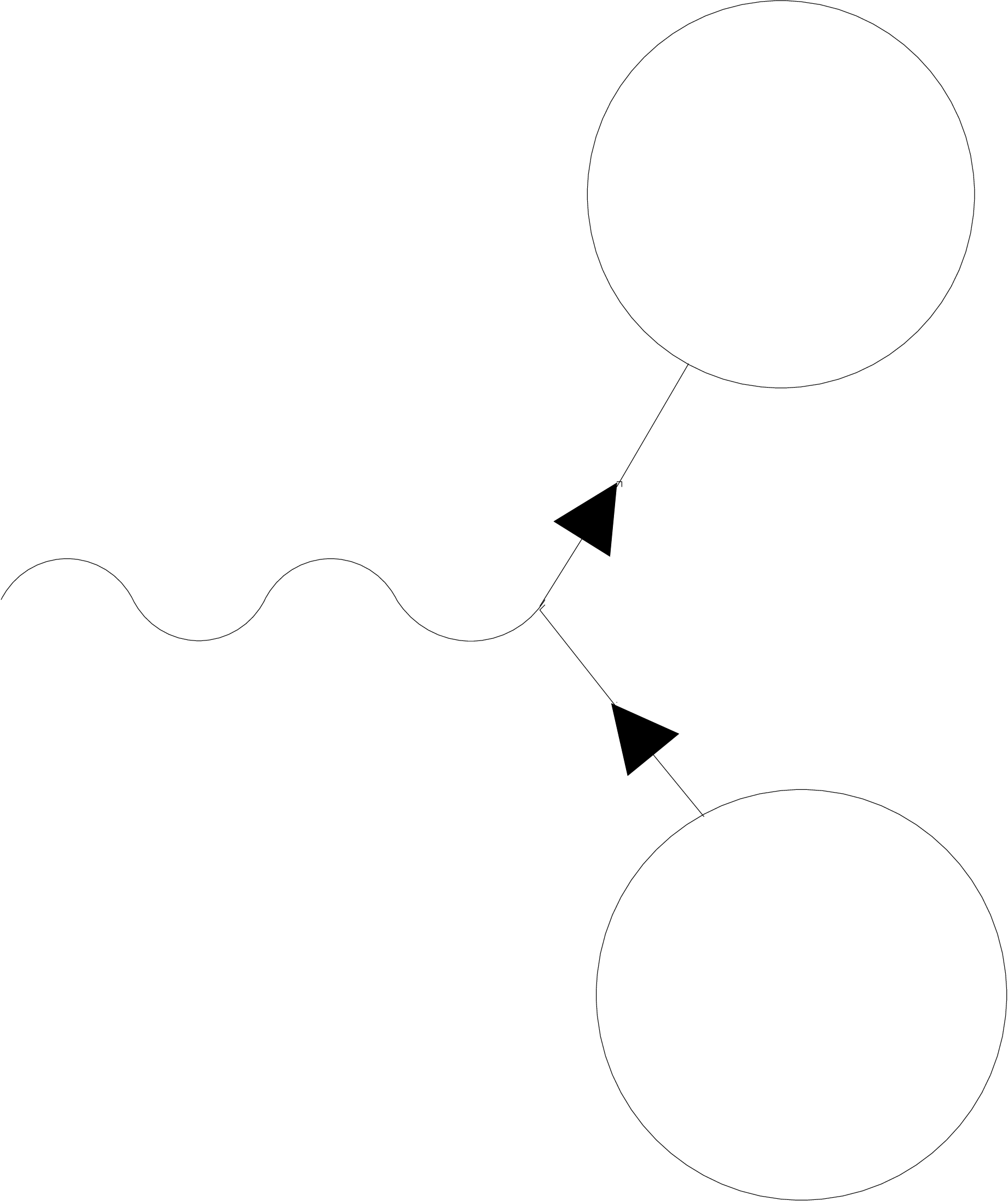}{-25}{1.8}{2}
\end{eqnarray}
\begin{eqnarray}
&& b^\mu(P)=\sum_{i=1}^n
\delta_{P=p_i} b^\mu(p_i) 
\nonumber 
\\
&&+ \sum_{P=P_1+P_2}
(ig)\Pi^\mu_\nu\; \psb(P_2)\gamma^\nu\psi(P_1)
\epsilon(P_1,P_2) 
\end{eqnarray}
\noindent
where 
\begin{eqnarray}
 b_\mu(P) \;=\; \plaata{bos1}{-5}{0.8}{0.5} \;\;\;\;
   \psi(P)  \;=\; \plaata{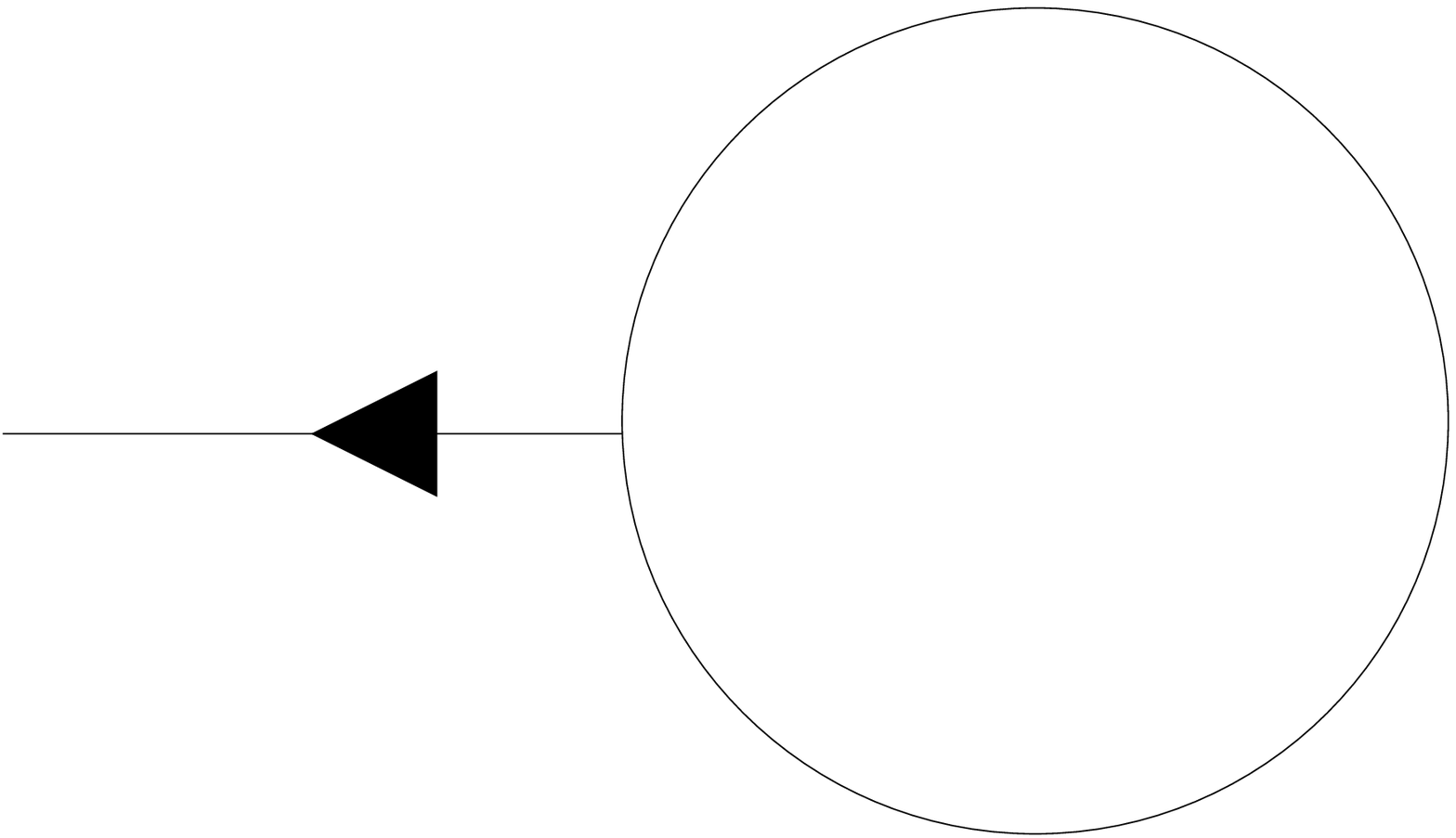}{-6}{0.8}{0.6}\; \;\;\;
   \psb(P)  \;=\; \plaata{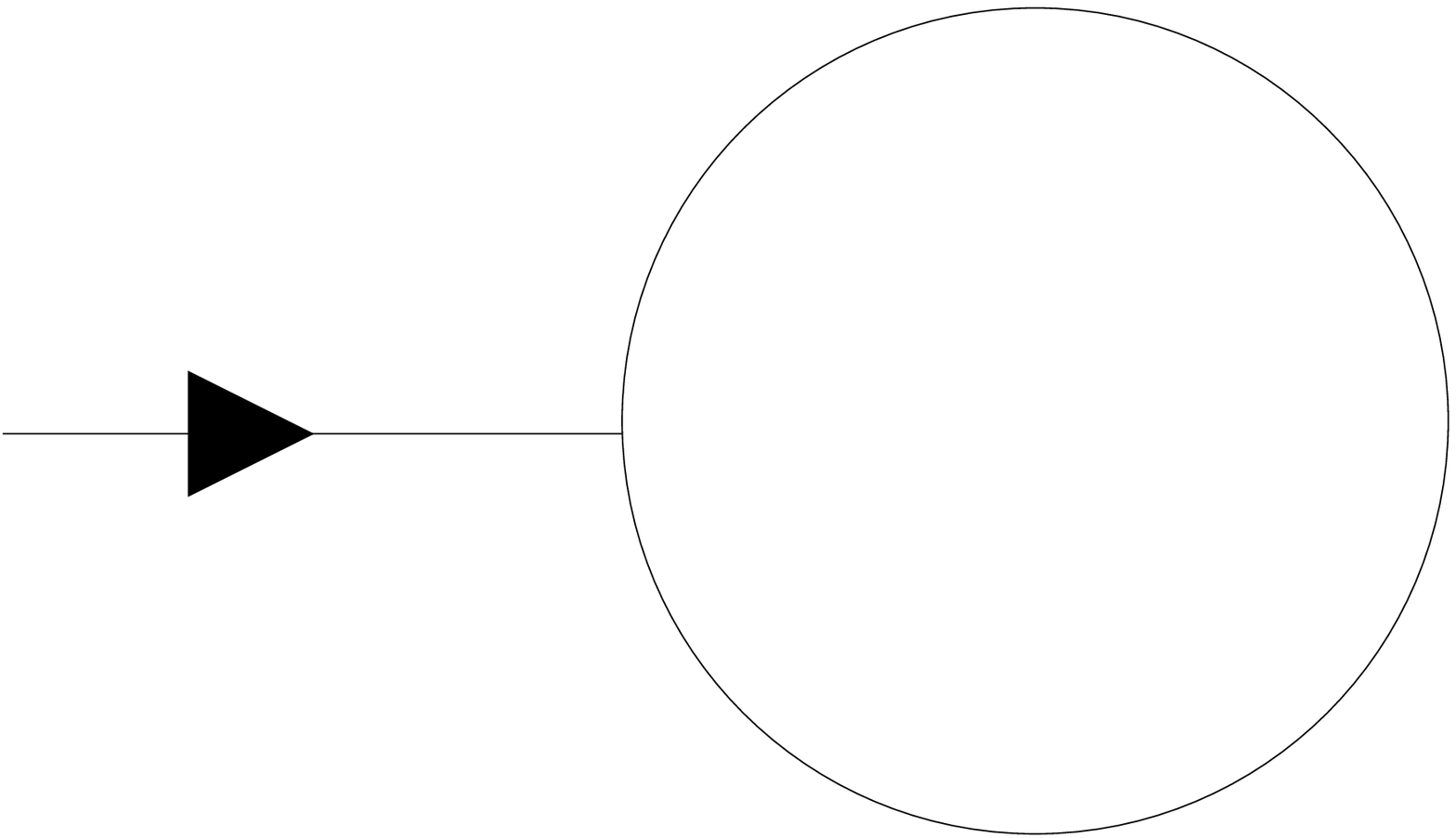}{-6}{0.8}{0.6} 
\end{eqnarray}
\noindent
describes a generic $n$-point Green function with one outgoing photon, 
fermion or anti\-fermion leg, respectively, carrying momentum $P$. 
$\Pi_{\mu\nu}$ stands for the boson propagator and $\epsilon$ takes into 
account the sign due to fermion anti-symmetrization.

\noindent
Technical details about the implementation of the algorithm for the
electroweak interactions can be found in \cite{Kanaki:2000ey}. 

\noindent
For QCD amplitudes, colour representation and summation play  an important
role. Usually, for the $n$-gluon amplitude the well known colour decomposition 
is used 
\begin{eqnarray}
{\cal M}= 2 i g^{n-2} \sum_{P(2,\ldots,n)} {\mathrm Tr}(t^{a_1}\ldots t^{a_n})
{\cal C}(1,\ldots,n)
\end{eqnarray}
\noindent
For processes involving quarks a similar expression may be derived. For further details,
the reader is referred to the vast literature on the subject~\cite{colourdecomposition}.
Methods for calculating the $C$-functions have been developed~\cite{gielethesis}, 
including some recent ones, 
more suitable for multiparticle processes~\cite{jeti,alphaqcd}. 
One of the most interesting aspects of this decomposition is the fact that 
the $C$-functions satisfy certain useful properties, such as gauge 
invariance and cyclic symmetry. Nevertheless, the computational complexity is 
rather high and the evaluation of the squared colour matrix a rather 
complicated task~\cite{kuijfthesis}.
 
\noindent
In {\tt HELAC} a novel approach has been considered. It is based on the colour 
connection (or colour flow) representation of the interaction vertices, where 
the explicit reference to the colour has been avoided, as is also the case 
in the usual colour decomposition. The advantage,  however,
is that the colour factors aquire a much simpler form, which moreover holds 
for gluon as well as for quark amplitudes, leading to a unified approach 
{\it for any tree-order process involving any number of coloured partons}. 
Moreover, the unweighting procedure is significantly facilitated, since the usual 
information on colour connections, needed by the parton shower Monte Carlo, is 
automatically available, without any further calculation. The colour factor 
is universally given by
\begin{eqnarray}
{\cal F}_I=\delta_{1\sigma_I(1)}\delta_{2\sigma_I(2)}
\ldots\delta_{n\sigma_I(n)}\,,
\end{eqnarray}
\noindent
whereas the colour matrix, defined as
\begin{eqnarray}
{\cal M}_{IJ} = \sum_{\mbox{\scriptsize  colours}} {\cal F}_I {\cal F}_J^\dagger
\end{eqnarray}
\noindent
with the summation running over all colours, $1\ldots N_c$, has a very simple
representation:
\begin{eqnarray}
{\cal M}_{IJ} = N_c^{m(\sigma_I,\sigma_J)}\,.
\end{eqnarray}
\noindent
Here, $1\le m(\sigma_I,\sigma_J)\le n$ counts the number of cycles made by the 
elements of the permutations $\sigma_I$ and $\sigma_J$. Details can be found
in ref.~\cite{cpp2001}.

\subsection{Phase-space integration: {\tt PHEGAS}}

\noindent
The study of multiparticle processes, such as six-fermion 
production in $e^+e^-$, requires efficient phase-space Monte Carlo generators.
The reason is that the squared amplitude, being a complicated function
of the kinematical variables, exhibits strong variations in specific regions and/or 
directions of the phase space, lowering in a substantial way the speed and the 
efficiency of the Monte Carlo integration. A well known way out of this problem relies on  
algorithms characterized by two main ingredients:

\begin{enumerate}
\item The construction of appropriate mappings of the phase space 
parametrization, in such a way that the main variation of the integrand
can be described by a set of almost uncorrelated  variables, and
\item A self-adaptation procedure that reshapes the generated phase-space
density in order to be as close as possible to the integrand.
\end{enumerate}

\noindent
In order to construct appropriate mappings, it is important to note that the integrand, 
i.e. the squared amplitude, has a well-defined representation in terms of Feynman diagrams. 
It is therefore natural to associate to each Feynman diagram a phase-space mapping that 
parametrizes the leading variation coming from it. In {\tt PHEGAS}, information from 
{\tt HELAC} is used to automatically construct a representation of all Feynman graphs 
contributing to the given process. The subset of Feynman graphs that results in a 
different 
phase-space parametrization is then used as kinematical mappings, called channels, 
to perform the Monte Carlo integration. Details can be found in~\cite{Papadopoulos:2000tt}.

\noindent
Since in six- and eight-fermion production a large number of kinematical channels 
contribute, typically of the order of $10^2$ to $10^4$, the optimization is also used 
to reduce their number. This is based on the fact that many of the channels exhibit an 
important correlation that renders them practically useless as separate channels.
The reduction in the number of channels achieved by this optimization is generally
important, resulting in a very efficient and rapid integration. 

\noindent
The main points can be summarized as follows:
\begin{itemize}
\item The algorithm exhibits a computational cost that grows like $\sim 3^n$, in contrast 
to the  $n!$ growth of the Feynman graph approach. Therefore there is no severe 
limitation in computing many-particle amplitudes (up to at least 12 external).
\item  All electroweak vertices in both the Feynman and the Unitary gauge
have been included, allowing highly non-trivial checks to be performed. The
QCD interactions have been implemented in the colour-connection
representation, allowing also a fast unweighting procedure. The decay width of
unstable particles is introduced in the fixed-width and complex-mass schemes. 
Any process with any type of Standard Model particle can be
reliably computed.
\item  Special features include also the possibility to use higher precision
floating point arithmetic, allowing full control over all possible
phase-space regions. Speeding up techniques, for helicity Monte Carlo treatment and
large $N_c$ estimates, are also available.
\item Incorporation of higher order corrections (currently available 
Fermion-Loop corrections up to three-point functions)
and the introduction of the Minimal Supersymmetric
Standard Model are in progress.
\end{itemize}

\section{The program {\tt AMEGIC++}}\label{manAMEGIC}

\noindent
{\tt AMEGIC++}, acronym for (A Matrix Element Generator in C++), 
is a multi-purpose parton-level generator written in C++. It provides 
a convenient tool for the calculation of cross sections for scattering 
processes at the tree level in the framework of the SM and the MSSM. 
Recently the code was extended to cover processes in the ADD 
model of large extra dimensions as well \cite{Gleisberg:2003ue}.
The program can also be used to generate single events and it is one
of the modules for the new complete event simulation framework 
{\tt SHERPA} \cite{Gleisberg:2003}. As such, the single events of {\tt AMEGIC++} 
can be handed over to the parton shower module {\tt APACIC++} 
\cite{Kuhn:2000dk} with the help of a new method that is correct at the 
next-to-leading logarithmic accuracy \cite{Catani:2001cc} and are thus 
linked correctly to fragmentation.  

\begin{figure}[h]
\begin{tabular}{cc}
\includegraphics[width=4cm]{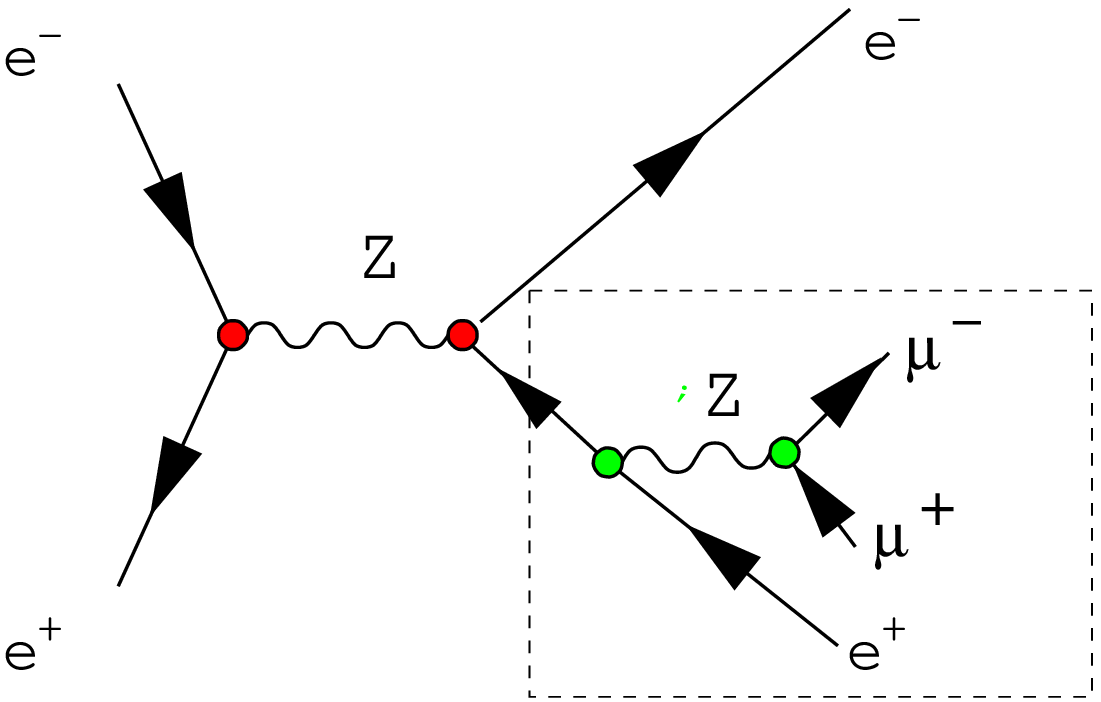} &
\includegraphics[width=4cm]{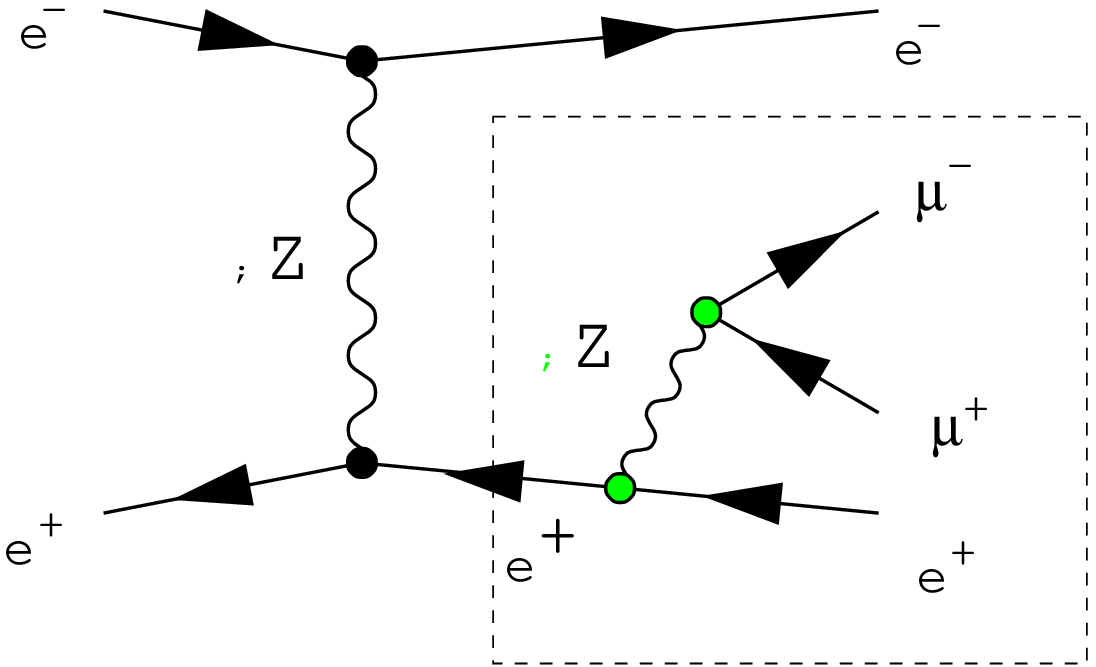} 
\end{tabular}
\caption{\label{superamp}
         Factoring out common pieces of amplitudes with identical colour 
         structure. In the example above, the parts within the boxes are 
         identical, hence the two amplitudes can be added and the terms
         inside the box can be factored out. }
\end{figure}

\noindent
In {\tt AMEGIC++}, full sets of Feynman diagrams are constructed
automatically and are translated by the program into helicity 
amplitudes in a formalism similar to the one in \cite{Kleiss:1985yh}. 
The colour structure of each diagram is represented as a word string, 
the emerging structures are grouped into sets of amplitudes with 
identical, common colour structure. Based on them, a matrix of colour
factors between amplitudes is calculated using the ordinary $SU(3)$
algebra. A number of refinements of the helicity method has been 
implemented within the code as well. First of all, the algorithm 
presented in \cite{Denner:1992vz} fixes the relative signs of amplitudes 
when Majorana fermions are present. Furthermore, explicit polarizations 
for massive or massless external spin-1 bosons are enabled, allowing us to 
consider polarized cross sections. Similar considerations help to
replace numerators of spin-1 propagators by summing over suitably
defined polarizations for off-shell particles disentangling nested 
Lorentz structures emerging for amplitudes with many internal
spin-1 bosons. As a result, {\tt AMEGIC++} needs only quite a limited
set of building blocks to construct all helicity amplitudes. Internally,
they are represented as word strings employing some knowledge-\-storing 
mechanism that ensures that all building blocks have to be evaluated
only once for each call of the full matrix element. With the help of internal 
methods these word strings are further simplified. Furthermore,
another massive gain in efficiency has been achieved by summing
amplitudes with identical colour structure and by
algorithms for finding common factors. This is exemplified in Fig.
\ref{superamp}. Having performed these manipulations, the resulting 
helicity amplitudes are stored in library files.

\noindent
There are a number of prescriptions to treat unstable particles. At the 
moment, {\tt AMEGIC++} supports the fixed-width scheme (FWS) and the 
complex-mass scheme (CMS). \\
Defining the complex mass parameters of the 
electroweak gauge bosons, the Higgs boson and the top quark in terms of 
the real masses and the constant widths through
\begin{eqnarray}
&& M_V^2 = m_V^2 - i\Gamma_V\,m_V\,,\quad V = W,Z\nonumber\\
&& M_H^2 = m_H^2 - i\Gamma_H\,m_H\,,\quad M_t = m_t - i\Gamma_t/2\,,
\end{eqnarray}
the corresponding propagators can be written as 
\begin{eqnarray}
&&D^{\mu\nu}_F(q) = \frac{-g^{\mu\nu} +
q^{\mu}q^\nu/M_V^2}{q^2-M_V^2}\,,\quad D_F(q) =
\frac{1}{q^2-M_H^2}\,,\nonumber\\ 
&&S_F(q) = \frac{\qslash+M_t}{q^2-M_t^2}\,.
\end{eqnarray}
In the FWS, the electroweak mixing angle is defined according to 
\begin{eqnarray}
\sin^2\theta_W = 1-\frac{m_W^2}{m_Z^2}\,. 
\end{eqnarray}
It is kept real. For the case of the gauge-invariant CMS, the real
gauge-boson masses have to be replaced by their complex counterparts
and this parameter is therefore complex as well.

\noindent
Within {\tt AMEGIC++} the Yukawa couplings of fermions to the Higgs boson
and their kinematical masses are decoupled. This allows us to study, for example,
the production of Higgs bosons and their decay into $b$-quarks, even in 
those cases where the user prefers to neglect the influence of the $b$-mass 
on both the phase space and the helicity structure.

\begin{figure}[h]
\begin{tabular}{cc}
\includegraphics[width=4cm]{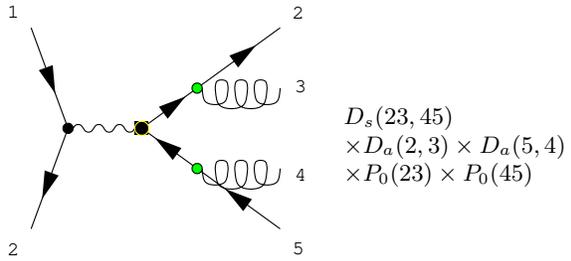} &
\begin{minipage}[t]{4cm}
\vspace*{-2cm}
$\begin{array}{l}
 D_s(23,45) \\
 \times D_a(2,3)\times D_a(5,4)\\
 \times P_0(23)\times P_0(45)
 \end{array}
$
\end{minipage}
\end{tabular}
\caption{\label{exchan}
         Translation of a Feynman diagram into a phase-space parametrization.
         $D_{s,a}$ denote symmetric or asymmetric decays ; the latter ones 
         reproduce the typical feature of collinear emission of particles
         notorious for gauge theories with massless spin-1 bosons. The
         propagator terms for massless particles $P_0$ peak at  the minimal
         allowed invariant mass.}
\end{figure}

\noindent
For the integration over the phase space of the outgoing particles,
{\tt AMEGIC++} employs an adaptive multichannel method 
\cite{Berends:1994pv}. Similar to their implementation, generic elements 
for phase-space mappings such as propagator-like structures
are provided. The individual Feynman diagrams are analyzed individually 
and one or more suitable phase-space parametrizations for each diagram
are automatically created and stored in library files. As an example,
consider Fig. \ref{exchan}, which exhibits a diagram and its translation 
into propagator- and decay-parametrizations. These files, both 
for the amplitudes and the phase-space parametrizations, are compiled
and linked to the code before the actual integration starts.

\noindent
For users of {\tt AMEGIC++} only very little intervention is needed.
Having specified the process(es), the model framework and its parameters,
a first ``initialization'' run of the code results in the creation of 
library files. After their compilation, a second, ``production'' run
will generate the results without any further manipulation.

\section{Numerical results}\label{numResults}
\subsection{Input parameters and phase-space cuts}
\noindent
The SM parameters are given in the $G_{\mu}$ scheme:  
\begin{eqnarray}
&&m_W = 80.419\;{\rm GeV}\,,\quad \Gamma_W = 2.12\;{\rm GeV,}\nonumber\\
&&m_Z = 91.1882\;{\rm GeV}\,,\quad \Gamma_Z = 2.4952\; {\rm GeV,}\nonumber\\
&&G_{\mu} = 1.16639 \times 10^{-5}\; {\rm GeV}^{-2},\nonumber\\
&&\sin^2\theta_W = 1 - m^2_W/m^2_Z,\nonumber\\  
&&\alpha_s = 0.0925(0.0891)\;\;{\rm at\;360(500)\;GeV}.
\end{eqnarray}
\noindent
The electromagnetic coupling is derived from the Fermi constant $G_{\mu}$ 
according to 
\begin{eqnarray}
&&\alpha_{\rm em} = \frac{\sqrt{2}\,G_{\mu}\,M^2_W\,\sin^2\theta_W}{\pi}. 
\end{eqnarray}
\noindent
The mass of the Higgs boson is assumed to be $M_H = 130$ GeV and its associated 
SM tree-level width is $\Gamma_H = 0.00429$ GeV. For this Higgs boson mass its 
branching ratios $H \to b\,\bar b$ and $H \to W^{\p}\,W^{\m} \to 4f$ are of the 
same order and therefore both decay channels signify its occurrence as an intermediate
state. For the massive fermions, the following masses have been used:
\begin{eqnarray}
&&m_{\mu} = 105.6583\;{\rm MeV,} \quad m_{\tau} = 1.777\;{\rm GeV,} \nonumber\\
&&m_u = 5\; {\rm MeV,} \quad m_d = 10\; {\rm MeV,}\nonumber\\
&&m_s = 200\;{\rm MeV,} \quad m_c = 1.3\; {\rm GeV,}\nonumber\\
&&m_b = 4.8\;{\rm GeV,}\nonumber\\
&&m_t = 174.3\; {\rm GeV,} \quad \Gamma_t = 1.6\; {\rm GeV.} 
\end{eqnarray}
\noindent
The constant widths of the electroweak gauge bosons, the Higgs boson
and the top quark are introduced via the fixed-width scheme as defined in 
Sect.~\ref{manAMEGIC}. CKM  mixing of the quark generations and the coupling 
of the Higgs boson to the very light fermion flavors ($e$, $u$, $d$)
is neglected.  

\noindent
Concerning the phase-space integration, the following cuts are applied
on the external particles:  
\begin{eqnarray}
&&\theta(l,{\rm beam}) > 5^\circ\;,\quad \theta(l,l') > 5^\circ\;,\quad E_l > 10\;{\rm GeV,}\nonumber\\
&&\theta(q,{\rm beam}) > 5^\circ\;,\quad \theta(l,q) > 5^\circ\;,\quad
E_q > 10\;{\rm GeV,}\nonumber\\
&&m(q,q') > 10\;{\rm GeV}\;,\label{cutsJegerlehner} 
\end{eqnarray}
\noindent
where $\theta(i,j)$ specifies the angle between the particles $i$ and
$j$ in the center-of-mass frame, and $l$, $q$ and beam
denote charged leptons, quarks or gluons and the beam electrons or
positrons, respectively. The invariant mass of a jet pair $qq'$ is
denoted by $m(q,q')$.  

\noindent
All results presented here are obtained using $10^6$ 
points (before cuts); 
statistical errors
of the Monte Carlo integrations, i.e. one standard deviation, are given in 
parentheses.  
\subsection{Results}
\begin{table}[h] 
\centerline{Top-quark channels}
{\small
\begin{center}
\begin{tabular}{lccc} 
\hline\noalign{\smallskip}      
Final state          & QCD & {\tt AMEGIC++} [fb] & {\tt
HELAC} [fb]\\
\noalign{\smallskip}\hline\noalign{\smallskip}
$b\bar b u \bar dd \bar u$ 
                     & yes & 32.90(15) & 33.05(14) \\
                     & yes & 49.74(21) & 50.20(13)\\    
                     & no & 32.22(34)  & 32.12(19)\\
                     & no & 49.42(44)  & 50.55(26)\\
$b\bar b u \bar ug g$ 
                     & -- & 11.23(10) & 11.136(41)\\
                     & -- & 9.11(13)  & 8.832(43)\\
$b\bar b g gg g$ 
                     & -- & 18.82(13) & 18.79(11)\\
                     & -- & 24.09(18) & 23.80(17)\\
$b\bar b u \bar d e^{\m} \bar \nu_e$ 
                     & yes & 11.460(36) & 11.488(15)\\
                     & yes & 17.486(66) & 17.492(41)\\
                     & no & 11.312(37)  & 11.394(18)\\
                     & no & 17.366(68)  & 17.353(31)\\
$b\bar b e^{\p} \nu_e e^{\m} \bar \nu_e$ 
                     & -- & 3.902(31) & 3.885(7)\\
                     & -- & 5.954(55) & 5.963(11)\\
$b\bar b e^{\p} \nu_e \mu^{\m} \bar \nu_{\mu}$ 
                     & -- & 3.847(15) & 3.848(7)\\
                     & -- & 5.865(24) & 5.868(10)\\
$b\bar b \mu^{\p} \nu_{\mu} \mu^{\m} \bar \nu_{\mu}$ 
                     & -- & 3.808(16) & 3.861(19)\\
                     & -- & 5.840(30) & 5.839(12)\\
\noalign{\smallskip}\hline
\end{tabular}
\end{center}
}
\caption{\label{top}
         The cross sections for possible signals and backgrounds 
         of top quark pair production in $e^+\,e^-$ annihilation. 
         All results in fb for $\sqrt{s} = 360$ GeV (first row) and
         $\sqrt{s} = 500$ GeV (second row).}
\end{table}
First of all, processes have been considered that serve as signals or backgrounds for
the production and decay of top pairs, Table \ref{top}. Since the branching ratio 
is practically 100\% for the decay of top quarks into bottom quarks and a $W$ 
($t\to bW^+$, $\bar t\to \bar bW^-$), all modes considered include a pair of bottom quarks. 
In cases involving a mixture of top production and decay and pure QCD diagrams,
 the relative 
importance of the different contributions to the total cross section has been estimated by 
switching on and off the QCD coupling constant. In both cases (the fully hadronic mode 
$b\bar bu\bar ud\bar d$ and the semileptonic mode $b\bar bu\bar de^-\bar\nu_e$) the top 
contribution is by far the dominating channel; the difference of taking into account the
QCD contributions or neglecting them is of the order of 2-3\%. Also, the total cross section 
of the fully hadronic channel is substantially larger than the cross section of any other 
individual $b\bar b$+4 jets mode.

\begin{table}[h] 
\centerline{Vector fusion with Higgs exchange}
{\small
\begin{center}
\begin{tabular}{lccc}  
\hline\noalign{\smallskip}      
Final state              & QCD & {\tt AMEGIC++} [fb] &
{\tt HELAC} [fb]\\
\noalign{\smallskip}\hline\noalign{\smallskip}
$e^{\m}e^{\p}u \bar ud \bar d$ 
                     & yes & 0.6842(85)&0.6858(31)\\
                     & yes & 1.237(15)&1.265(5)\\
                     & no  & 0.6453(62)&0.6527(35)\\
                     & no  & 1.206(14)&1.2394(75)\\
$e^{\m}e^{\p}u \bar ue^{\m} e^{\p}$ 
                     & -- & 6.06(36)e-03 &6.113(87)e-03\\
                     & -- & 6.58(23)e-03 &6.614(80)e-03\\
$e^{\m}e^{\p}u \bar u\mu^{\m} \mu^{\p}$  
                     & -- & 9.24(12)e-03 &9.04(11)e-03\\
                     & -- & 9.25(17)e-03 &9.145(74)e-03\\
$\nu_e\bar \nu_eu \bar dd \bar u$ 
                     & yes & 1.15(3)&1.176(6)\\
                     & yes & 2.36(7)&2.432(12)\\
                     & no  & 1.14(3)&1.134(5)\\
                     & no  & 2.35(7)&2.429(13)\\
$\nu_e\bar \nu_eu \bar de^{\m} \bar \nu_e$ 
                     & -- & 0.426(11) &0.4309(48)\\ 
                     & -- & 0.916(30) & 0.9121(48)\\
$\nu_e\bar \nu_eu \bar d\mu^{\m} \bar \nu_{\mu}$ 
                     & -- & 0.425(12) &0.4221(30)\\
                     & -- & 0.878(27)&0.8888(47)\\
\noalign{\smallskip}\hline
\end{tabular}
\end{center}
}
\caption{\label{bosons1}
         The cross sections for different $e^+\,e^- \to 6f$ final states
         corresponding to the Higgs production via vector-boson 
         fusion signal. All results in fb
         for $\sqrt{s} = 360$ GeV (first row) and
         $\sqrt{s} = 500$ GeV (second row).}
%
\centerline{Vector fusion without Higgs exchange}
{\small
\begin{center}
\begin{tabular}{lccc} 
\hline\noalign{\smallskip}      
Final state              & QCD & {\tt AMEGIC++} [fb] &
{\tt HELAC} [fb]\\
\noalign{\smallskip}\hline\noalign{\smallskip}
$e^{\m}e^{\p}u \bar ud \bar d$ 
                     & yes & 0.4838(50) &0.4842(25)\\
                     & yes & 1.0514(97) &1.0445(51)\\
                     & no  & 0.4502(31)  &0.4524(23)\\
                     & no  & 1.0239(79)  &1.0227(43)\\
$e^{\m}e^{\p}u \bar ue^{\m} e^{\p}$ 
                     & -- & 3.757(98)e-03 &3.577(43)e-03\\
                     & -- & 4.082(56)e-03 &4.214(46)e-03\\
$e^{\m}e^{\p}u \bar u\mu^{\m} \mu^{\p}$ 
                     & -- & 5.201(61)e-03 &5.119(70)e-03\\
                     & -- & 5.805(67)e-03 &5.828(49)e-03\\
$\nu_e\bar \nu_eu \bar dd \bar u$ 
                     & yes & 0.15007(53) &0.15070(64)\\
                     & yes & 0.4755(21)  &0.4711(24)\\
                     & no & 0.12828(42)  &0.12793(55)\\
                     & no & 0.4417(19)   &0.4398(21)\\
$\nu_e\bar \nu_eu \bar de^{\m} \bar \nu_e$ 
                     & -- & 0.04546(13) &0.04564(19)\\
                     & -- & 0.16033(63) &0.16011(78)\\
$\nu_e\bar \nu_eu \bar d\mu^{\m} \bar \nu_{\mu}$ 
                     & -- & 0.04230(12)  &0.04180(16)\\
                     & -- & 0.14383(53) &0.14439(65)\\
\noalign{\smallskip}\hline
\end{tabular}
\end{center}
}
\caption{\label{bosons2}
         The backgrounds to Higgs production via vector boson fusion. 
         All contributions from intermediate Higgs bosons are neglected. 
         Cross sections are given in fb for $\sqrt{s} = 360$ GeV
         (first row) and $\sqrt{s} = 500$ GeV (second row).}
\end{table}

For the QCD contributions, a similar pattern arises also in the vector-boson 
fusion channels, 
cf. Tables \ref{bosons1} and \ref{bosons2}. These channels are characterized by either an
electron-positron or an electron-neutrino anti-neutrino pair in the final state, 
corresponding
to either $Z$ boson or to $W$ boson fusion processes, respectively. 
\begin{table}[h] 
\centerline{Higgs production through Higgs-strahlung}
{\small
\begin{center}
\begin{tabular}{lccc} 
\hline\noalign{\smallskip}              
Final state              & QCD & {\tt AMEGIC++} [fb]
&{\tt HELAC} [fb]\\
\noalign{\smallskip}\hline\noalign{\smallskip}
$\mu^{\m}\mu^{\p}\mu^{\m} \bar \nu_{\mu}e^{\m} \bar \nu_e$ 
                     & -- & 0.03244(27) &0.03210(15)\\
                     & -- & 0.03747(29) &0.03749(32)\\
$\mu^{\m}\mu^{\p}u \bar de^{\m} \bar \nu_e$ 
                     & -- & 0.0924(8) &0.09306(46)\\
                     & -- & 0.1106(22) &0.10901(66)\\
$\mu^{\m}\mu^{\p}\mu^{\m}  \mu^{\p}e^{\m} e^{\p}$ 
                     & -- & 2.828(67)e-03 &2.923(52)e-03\\
                     & -- & 2.731(65)e-03 &2.691(42)e-03\\
$\mu^{\m}\mu^{\p}u \bar ud \bar d$ 
                     & yes & 0.2534(24) &0.2540(16)\\
                     & yes & 0.2634(22) &0.2642(15)\\
                     & no  & 0.2441(23)  &0.2471(15)\\
                     & no  & 0.2593(22)  &0.2589(14)\\
$\mu^{\m}\mu^{\p}u \bar uu \bar u$ 
                     & yes & 1.125(8)e-02   &1.135(22)e-02\\
                     & yes & 8.767(65)e-03  &8.978(58)e-03\\    
                     & no  & 7.929(57)e-03  &8.078(92)e-03\\
                     & no  & 6.098(35)e-03  &6.013(26)e-03\\
\noalign{\smallskip}\hline
\end{tabular}
\end{center}
}
\caption{\label{higgs1}
         The cross sections for different $e^+\,e^- \to 6f$ final states
         corresponding to the Higgs-strahlung signal. All
         results given in fb for $\sqrt{s} =
         360$ GeV (first row) and $\sqrt{s} = 500$ GeV (second row).}
%
\centerline{Backgrounds to Higgs-strahlung}
{\small
\begin{center}
\begin{tabular}{lccc} 
\hline\noalign{\smallskip}              
Final state              & QCD & {\tt AMEGIC++} [fb] &
{\tt HELAC} [fb]\\
\noalign{\smallskip}\hline\noalign{\smallskip}
$\mu^{\m}\mu^{\p}\mu^{\m} \bar \nu_{\mu}e^{\m} \bar \nu_e$ 
                     & -- & 0.01845(14) &0.01843(13)\\
                     & -- & 0.03054(23) &0.03092(19)\\
$\mu^{\m}\mu^{\p}u \bar de^{\m} \bar \nu_e$ 
                     & -- & 0.05284(57) &0.05209(33)\\
                     & -- & 0.08911(53) &0.08925(48)\\
$\mu^{\m}\mu^{\p}\mu^{\m}  \mu^{\p}e^{\m} e^{\p}$ 
                     & -- & 2.204(52)e-03 &2.346(49)e-03\\
                     & -- & 2.280(66)e-03 &2.277(62)e-03\\
$\mu^{\m}\mu^{\p}u \bar ud \bar d$ 
                     & yes & 0.1412(10) &0.1404(11)\\
                     & yes & 0.2092(12) &0.2075(13)\\   
                     & no  & 0.1358(20)  &0.1341(12)\\
                     & no  & 0.2040(12)  &0.2015(11)\\
$\mu^{\m}\mu^{\p}u \bar uu \bar u$ 
                     & yes & 5.937(24)e-03 &5.937(25)e-03\\
                     & yes & 6.134(29)e-03 &6.108(27)e-03\\     
                     & no  & 2.722(10)e-03  &2.710(11)e-03\\
                     & no  & 3.290(12)e-03  &3.303(12)e-03\\
\noalign{\smallskip}\hline
\end{tabular}
\end{center}
}
\caption{\label{higgs2}
         Background contributions to the Higgsstrahlungs signal for
         various $6f$ final states. All diagrams with intermediate
         Higgs bosons have been neglected. Cross sections are given 
         in fb for $\sqrt{s} = 360$ GeV (first row) and 
         $\sqrt{s} = 500$ GeV (second row).}
\end{table}
Again, switching on and off 
the QCD coupling constant gives rise to differences on the level of a few per cent. In contrast, 
taking into account the Higgs boson (Table \ref{bosons1}) which may be produced in the $s$-channel 
through the fusion of two $t$-channel vector bosons, or neglecting it (Table \ref{bosons2}) 
changes the total cross sections for all channels considered by a factor of 2 or larger. 
This is especially pronounced for channels that can be identified as $WW$-fusion channels with 
a semileptonic or fully hadronic decay of the $W$-pair produced by the Higgs decay 
(i.e. $\nu_e\bar \nu_eu \bar de^{\m} \bar \nu_e$ and 
$\nu_e\bar \nu_eu \bar d\mu^{\m} \bar \nu_{\mu}$, or $\nu_e\bar \nu_eu \bar dd \bar u$, 
respectively), where the cross sections are larger by one order of magnitude.

Another mode for Higgs production at an electron-positron collider is Higgs-strahlung, where
the Higgs boson is radiated off a $Z$-boson in the $s$-channel. In Table \ref{higgs1}, total cross 
sections for such modes are displayed, where the $Z$ boson decays into muons and the Higgs boson 
goes into four fermions through a pair of $W$ or $Z$ bosons. In Table \ref{higgs2}, identical 
total cross sections for the same final states, but neglecting the Higgs contribution, are
shown. In both cases, again, the size of the pure QCD contributions is found to be
small for most final states, i.e. of the order of few per cent. The only exception is for
a pair of muons and four identical quarks; there, the inclusion of QCD changes the results
by roughly 20\%, when the Higgs boson is taken into account, and by a factor of roughly 2
when its contribution is neglected. It is amusing to note that this relative factor of two
compares in size with the effect of including the Higgs boson itself. This, however, is true
only for the mode that 
can be imagined as $e^+e^-\to ZH\to ZZZ\to \mu^+\mu^-u\bar uu\bar u$.
In all other cases, as said before, inclusion of QCD has minor effects only; the Higgs boson
in contrast roughly doubles the total cross section in all the other channels.
\begin{table}[h] 
\centerline{Triple Higgs coupling}
{\small
\begin{center}
\begin{tabular}{lccc} 
\hline\noalign{\smallskip}              
Final state              & QCD & {\tt AMEGIC++} [fb] &
{\tt HELAC} [fb]\\
\noalign{\smallskip}\hline\noalign{\smallskip}
$\mu^{\m}\mu^{\p}b \bar bb \bar b$  
                     & yes & 2.560(26)e-02 &2.583(26)e-02\\
                     & yes & 3.096(60)e-02 &3.019(43)e-02\\
                     & no  & 1.711(55)e-02  &1.666(28)e-02\\
                     & no  & 2.34(12)e-02   &2.36(10)e-02\\
\noalign{\smallskip}\hline
\end{tabular}
\end{center}
}
\caption{\label{triple1} Cross sections for the process 
         $e^+\,e^- \to \mu^{\m}\mu^{\p}b \bar bb \bar b$.
         All results in fb for $\sqrt{s} =
         360$ GeV (first row) and $\sqrt{s} = 500$ GeV (second row).}
%
\centerline{Backgrounds to triple Higgs coupling}
{\small
\begin{center}
\begin{tabular}{lccc} 
\hline\noalign{\smallskip}      
Final state              & QCD & {\tt AMEGIC++} [fb] &
{\tt HELAC} [fb]\\
\noalign{\smallskip}\hline\noalign{\smallskip}
$\mu^{\m}\mu^{\p}b \bar bb \bar b$  
                     & yes & 7.002(32)e-03 & 7.044(22)e-03\\
                     & yes & 6.308(24)e-03 & 6.364(21)e-03\\
                     & no  & 2.955(11)e-03 & 2.972(12)e-03\\
                     & no  & 3.704(15)e-03 & 3.695(13)e-03\\
\noalign{\smallskip}\hline
\end{tabular}
\end{center}
}
\caption{\label{triple2} Cross sections for
         $e^+\,e^- \to \mu^{\m}\mu^{\p}b \bar bb \bar b$ with all
         contributions due to intermediate Higgs bosons left out. 
         All results in fb taken for $\sqrt{s} =
         360$ GeV (first row) and $\sqrt{s} = 500$ GeV (second row).}
\end{table}

One of the salient research goals at a potential linear collider operating at energies around
500 GeV is the determination of the Higgs potential. For this, the self-couplings of the Higgs bosons
have to be checked. In the framework of this publication, results are provided for the channel where
the Higgs bosons emerge in Higgs-strahlungs-like topologies and decay into a pair of bottom quarks.
This leads to final states $\mu^+\mu^-+4b$, where the muons mainly come from the $Z$ bosons.
Results for total cross sections for the process $e^+e^-\to \mu^+\mu^-+4b$, where contributions 
mediated by Higgs bosons have been included or neglected, are given in Tables \ref{triple1} and 
\ref{triple2}, respectively. From the results displayed one can read off that the inclusion of
intermediate Higgs bosons enhances the cross sections by a factor of three to four. Again, also 
the effect of QCD has been checked. For the process involving the intermediate Higgs bosons,
QCD leads to total cross sections that are larger by roughly 30\%-40\%, 
without the Higgs bosons,
QCD contributes on the level of factors of two to three.

\section{Summary of results}\label{Conclusion}
\begin{figure}[h]
\begin{center}
\includegraphics[width=8cm]{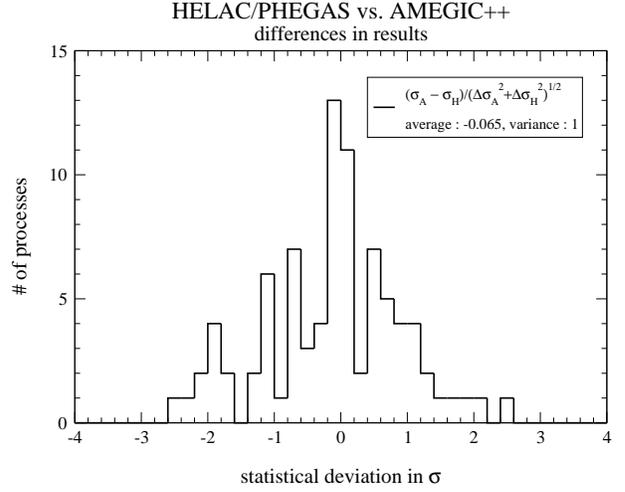}
\caption{The distribution of deviations $s^{(i)}$, given by Eq.\ref{deviationformula},
         for the eighty-six total cross sections $i$ presented in this paper. The average
         value is $\bar s = -0.065$, their variance is $\sigma_s \approx 1$\,.
         \label{deviations}}
\end{center}
\end{figure}

In the framework of this comparison, total cross sections for 86 different
processes involving six-particle final states have been obtained by the two 
multi-purpose matrix element generator packages {\tt HELAC/PHEGAS} and
{\tt AMEGIC++}. The integration over the multidimensional phase space of
the final states has been performed with Monte Carlo methods, and in all cases
one million MC points have been used. For nearly all cross sections the resulting
statistical error was significantly smaller than one per cent, roughly five per 
mille. There have been no significant differences between the two codes. To compare 
the results, for each process $i$ the deviation $s^{(i)}$ of the two
resulting cross sections $\sigma^{(i)}_{\rm H}$ and $\sigma^{(i)}_{\rm A}$ has been
calculated through
\begin{eqnarray}\label{deviationformula}
s^{(i)} = 
\frac{\sigma^{(i)}_{\rm A}-\sigma^{(i)}_{\rm H}}
     {\sqrt{\left(\Delta\sigma^{(i)}_{\rm A}\right)^2+
            \left(\Delta\sigma^{(i)}_{\rm H}\right)^2}}\,.
\end{eqnarray}
The distribution of the individual differences is depicted in Fig. \ref{deviations}.
The average deviation is $\bar s = -0.065$, the variance in their distribution is
$\sigma_s \approx 1$. The maximal difference between two cross sections is
smaller than three standard deviations, $s^{({\rm max.})}\approx 2.6$. The distribution
of differences follows roughly a Gaussian distribution. 

To summarize: Both packages, {\tt HELAC/PHEGAS} as well as {\tt AMEGIC++}, lead, with quite different 
methods, to consistent results for total cross sections for a large number of different
processes with six particles in the final state. This provides an independent check of the
precision level of the two codes, which can be considered as successfully tested.

\begin{acknowledgement}
{\it Acknowledgment:} The authors thank the Center for High Performance Computing Dresden (ZHR) for 
providing their resources and BMBF for financial support. The work of FK was supported by the EC 
5th Framework Programme under contract number HPMF-CT-2002-01663. CGP also acknowledges support 
from the EC project "Multi-particle Processes and Higher Order Corrections", HPMF-CT-2002-01622. 
SS wants to thank GSI Darmstadt for financial support. 
\end{acknowledgement}


\newpage
\begin{thebibliography}{99}
\bibitem{Kanaki:2000ey} A.~Kanaki and C.~G.~Papadopoulos,
                        Comput.\ Phys.\ Commun.\  {\bf 132} (2000) 306
                        [arXiv:hep-ph/0002082].
\bibitem{Papadopoulos:2000tt}
                        C.~G.~Papadopoulos,
                        Comput.\ Phys.\ Commun.\  {\bf 137} (2001) 247
                        [arXiv:hep-ph/0007335].

\bibitem{Krauss:2001iv} F.~Krauss, R.~Kuhn and G.~Soff,
                        JHEP {\bf 0202} (2002) 044
                        [arXiv:hep-ph/0109036].




\bibitem{Dittmaier:2002ap} S.~Dittmaier and M.~Roth,
                        Nucl.\ Phys.\ B {\bf 642} (2002) 307
                        [arXiv:hep-ph/0206070].
\bibitem{Kolodziej:2002ev} K.~Kolodziej,
                        arXiv:hep-ph/0210199.

\bibitem{Berends:1994pv} F.~A.~Berends, R.~Pittau and R.~Kleiss,
                        Nucl.\ Phys.\ B {\bf 424} (1994) 308
                        [arXiv:hep-ph/9404313];
                        R.~Kleiss and R.~Pittau,
                        Comput.\ Phys.\ Commun.\  {\bf 83} (1994) 141
                        [arXiv:hep-ph/9405257].
\bibitem{Gangemi:1998vc}
F.~Gangemi, G.~Montagna, M.~Moretti, O.~Nicrosini and F.~Piccinini,
Eur.\ Phys.\ J.\ C {\bf 9} (1999) 31
[arXiv:hep-ph/9811437];
F.~Gangemi, G.~Montagna, M.~Moretti, O.~Nicrosini and F.~Piccinini,
Nucl.\ Phys.\ B {\bf 559} (1999) 3
[arXiv:hep-ph/9905271];
F.~Gangemi,
arXiv:hep-ph/0002142.
%
\bibitem{Moretti:2001zz} M.~Moretti, T.~Ohl and J.~Reuter,
                        arXiv:hep-ph/0102195.
\bibitem{Kilian:2001qz} W.~Kilian,
                        LC-TOOL-2001-039.

\bibitem{Stelzer:1994ta} T.~Stelzer and W.~F.~Long,
                        Comput.\ Phys.\ Commun.\  {\bf 81} (1994) 357
                        [arXiv:hep-ph/9401258].
\bibitem{Maltoni:2002qb} F.~Maltoni and T.~Stelzer,
                        JHEP {\bf 0302} (2003) 027
                        [arXiv:hep-ph/0208156].


\bibitem{Caravaglios:1995cd} F.~Caravaglios and M.~Moretti,
                        Phys.\ Lett.\ B {\bf 358} (1995) 332
                        [arXiv:hep-ph/9507237];
                        Z.\ Phys.\ C {\bf 74} (1997) 291
                        [arXiv:hep-ph/9604316].
\bibitem{Ohl:1998jn} T.~Ohl,
                        Comput.\ Phys.\ Commun.\  {\bf 120} (1999) 13
                        [arXiv:hep-ph/9806432].
\bibitem{Murayama:1992gi} H.~Murayama, I.~Watanabe and K.~Hagiwara,
                        KEK-91-11.

\bibitem{Grunewald:2000ju} M.~W.~Grunewald {\it et al.},
                        arXiv:hep-ph/0005309.
\bibitem{Aguilar-Saavedra:2001rg} J.~A.~Aguilar-Saavedra {\it et al.}, 
                        [The ECFA/DESY LC Physics Working Group], 
                        hep-ph/0106315.
%
\bibitem{Dittmaier:2003sc} S.~Dittmaier, 
[arXiv:hep-ph/0308079].

\bibitem{colourdecomposition}
M.~L.~Mangano and S.~J.~Parke,
``Multi-parton amplitudes in gauge theories'',
Phys. Rep. {\bf 200}, (1991) 301-367, and references therein.
[arXiv:hep-ph/9212246].

\bibitem{gielethesis}
W.T.~Giele,
``Properties and calculations of multiparton processes'',
PhD thesis (University of Leiden) (1989).

\bibitem{jeti}
P.~Draggiotis, R.~H.~P.~Kleiss and C.~G.~Papadopoulos,
Phys.\ Lett.\ B {\bf 439} (1998) 157
[arXiv:hep-ph/9807207];\\
P.~D.~Draggiotis, R.~H.~P.~Kleiss and C.~G.~Papadopoulos,
Eur.\ Phys.\ J.\ C {\bf 24} (2002) 447
[arXiv:hep-ph/0202201].

\bibitem{alphaqcd}
F.~Caravaglios, M.~L.~Mangano, M.~Moretti and R.~Pittau,
``A new approach to multi-jet calculations in hadron collisions,''
Nucl.\ Phys.\  {\bf B539} (1999) 215
[hep-ph/9807570];
M.~L.~Mangano, M.~Moretti, F.~Piccinini, R.~Pittau and A.~D.~Polosa,
JHEP {\bf 0307} (2003) 001
[arXiv:hep-ph/0206293].


\bibitem{kuijfthesis}
J.G.M.~Kuijf,
``Multiparton production at hadron colliders'', PhD thesis
(University of Leiden) (1991)

\bibitem{cpp2001}
"HELAC-PHEGAS: automatic computation of helicity amplitudes and cross
sections" by A.~Kanaki and C.~G.~Papadopoulos,  Second CPP
Symposium-Computational Particle Physics , KEK Proceedings 2002-11,
Editor: Y.~Kurihara, August 2002, pp. 20-25.

\bibitem{Gleisberg:2003ue}
T.~Gleisberg, F.~Krauss, K.~T.~Matchev, A.~Sch{\"a}licke, 
S.~Schumann and G.~Soff,
JHEP {\bf 0309} (2003) 001
[arXiv:hep-ph/0306182].
\bibitem{Gleisberg:2003}
T.~Gleisberg, S.~H{\"o}che, F.~Krauss, A.~Sch{\"a}licke, 
S.~Schumann and J.~Winter,
CERN--TH/2003-284 and hep-ph/0311263.
\bibitem{Kuhn:2000dk} R.~Kuhn, F.~Krauss, B.~Ivanyi and G.~Soff,
                        Comput.\ Phys.\ Commun.\  {\bf 134} (2001) 223
                        [arXiv:hep-ph/0004270].
\bibitem{Catani:2001cc} S.~Catani, F.~Krauss, R.~Kuhn and B.~R.~Webber,
                        JHEP {\bf 0111} (2001) 063
                        [arXiv:hep-ph/0109231]; 
                        F.~Krauss,
                        JHEP {\bf 0208} (2002) 015
                        [arXiv:hep-ph/0205283].

\bibitem{Kleiss:1985yh} R.~Kleiss and W.~J.~Stirling,
                        Nucl.\ Phys.\ B {\bf 262} (1985) 235;
                        A.~Ballestrero, E.~Maina and S.~Moretti,
                        Nucl.\ Phys.\ B {\bf 415} (1994) 265
\bibitem{Denner:1992vz}
A.~Denner, H.~Eck, O.~Hahn and J.~Kublbeck,
Nucl.\ Phys.\ B {\bf 387} (1992) 467.

\end{thebibliography}
\end{document}